\documentclass{mem}
\usepackage{natbib}
\usepackage{txfonts}
\usepackage{balance}
\usepackage{graphicx}
\usepackage{txfonts}
\usepackage[a4paper,breaklinks,dvipdfm]{hyperref}
\bibpunct{(}{)}{;}{a}{}{,} 

\idline{1}{1}

\begin{document}
\def\teff{$T\rm_{eff }$}
\def\kms{$\mathrm {km s}^{-1}$}
\def\fcgs{erg cm$^{-2}$ s$^{-1}$}
\def\lxcgs{erg s$^{-1}$}

\title{CV surveys with eROSITA}

\author{A. Schwope
\inst{1}
}
\offprints{A. Schwope}
\institute{
Leibniz-Institute f\"{u}r Astrophysik Potsdam (AIP), An der Sternwarte 16,
14482 Potsdam, Germany  
\email{aschwope@aip.de}
}

\authorrunning{Schwope}

\titlerunning{CV surveys with eROSITA}

\abstract{eROSITA will perform the most sensitive X-ray all-sky ever in the
  energy range $0.3 - 10$ keV. It will likely uncover several $10^4$ compact
  binaries, most of them  will be cataclysmic variable stars. After a brief
  introduction to eROSITA we will discuss  the source content of the eROSITA
  surveys, the expected number of CVs and possible strategies  for the optical
  follow-up.   
\keywords{Stars: cataclysmic variables --
Stars: X-rays -- surveys}
}
\maketitle{}

\section{Introduction}
eROSITA (extended ROentgen Survey with an Imaging Telescope Array) is an X-ray 
telescope array built by a German consortium, to be mounted on a Fregat
booster and launched with a Russian Zenit rocket  into an L2 halo orbit of the
Sun-Earth system in 
late 2013 \citep{predehl+10, pavlinsky+09}.  During the first four years of
its mission, the eROSITA telescopes will scan the sky in great circles with
a scanning speed corresponding to one full circle being completed every four
hours. The scanning axis is either pointed  directly towards the sun or
a few degrees away from it. As the satellite  moves around the
sun, the plane of the scans is thus advanced by about one degree  per day,
resulting in a full coverage of the sky every half year.  

Hence eROSITA will perform eight independent all-sky surveys
in the energy range $0.3-10$\,keV each lasting half a year. After four years
the sky area around the poles of the ecliptic will be exposed for about 20 ks, 
the average exposure per survey is 240\,s (eRASS:1) and for the full four year
survey about 2000\,s (eRASS:8). 
The average sensitivity for the detection of point sources in eRASS:8
will be $1.5 \times 10^{-14}$\,\fcgs\ and $2\times 10^{-13}$\,\fcgs\ in the
soft ($0.5 - 2.0$\,keV) and hard ($2 - 10$\,keV)  energy bands,
respectively. Around  the poles of the ecliptic the limiting fluxes will be
$4\times10^{-15}$\,\fcgs\ (soft) and $5\times10^{-14}$\,\fcgs (hard).  

The X-ray optics will have an on-axis angular resolution of 15\arcsec\ (HEW,
half energy width), due to off-axis blurring the average survey resolution will be lower
and is expected to be below 30 arc sec. Individual source positions will have a
remaining positional inaccuracy of estimated $2-3$ arc sec.  

Each of the seven mirror modules has its own camera in its focus, each
equipped with a CCD-module and a  processing electronics. The eROSITA-CCDs
have $384\times 384$ pixels which correspond to an image area of 28.8 mm
$\times$ 28.8 mm, respectively, for a field of view of 1.03\degr\ diameter.
The 384 channels are read out in parallel and the nominal integration time per
CCD frame will be 50 ms.  

Given the technical specifications, eROSITA can be very much compared to
XMM-Newton with EPIC pn, it has however a factor 5 times larger grasp between
0.3 and 2 keV (effective area times FoV), making it a prime
survey instrument. During its four-year all-sky survey it shall discover about
$10^5$ clusters of galaxies to study the growth of structure and constrain the
parameters of Dark Energy as the prime mission goal. Based on the known
number-flux relations of AGN \citep[e.g.][]{gilli}) one may expect the detection
of about $3\times 10^6$\,AGN of all kind up to redshift 6. 
The eROSITA X-ray sky will also be 
populated by about 300,000 coronal emitters. The huge stellar samples built
from the large number of detections will help to disentangle
the X-ray emitting populations and determine the shape of
age-metalicity-activity relations.

Finally, compact galactic (and extragalactic) objects will be detected in
large number and in the following sections the expectations are outlined. 

\section{CV forecast for eROSITA}
The current version of the Catalogue of Cataclysmic Variables, Low-Mass X-ray
Binaries and related objects \citep[RKCat Edition 7.16][]{rkcat}) has entries
for 926 CVs. Despite this comparatively large number, flux-limited samples that
could be used to derive the space density of the CV population are
comparatively small. The most comprehensive X-ray selected complete sample was
collected as part of the ROSAT Bright Survey (RBS), an optical identification
program of the more than 2000 brightest X-ray sources at high-galactic
latitude \citep{schwope+00,schwope+02} und comprised 16 non-magnetic CVs,
among them 5 new discoveries. The RBS sample of non-magnetic CVs was 
analyzed jointly with the CV sample from the ROSAT North Ecliptic Pole survey
\citep{gioia+03} by \cite{pretorius+07} and more recently by
\cite{pretorius_knigge11}. The NEP sample is small, it comprises just 4
objects. The NEP reached a flux limit comparable to eRASS:8
over a sky area of only 81 square degrees; the RBS on the other hand 
reached $\sim 1\times 10^{-12}$\,\fcgs\ over 20000 square degrees. 

The joint analysis of NEP and RBS non-magnetic CVs revealed a
luminosity function in the range 
$\log{\rm L_X} = 29.7 - 31.8$\,\lxcgs. Integration of the observed luminosity 
function gave a mid-plane space density of  
$\rho_0 \simeq 6\times 10^{-6}$\,pc$^{-3}$. The error in this quantity is of
order a factor 2. Perhaps the biggest source of uncertainty therein is the
often badly constrained distance to individual CVs. Because the
number of CVs in those X-ray flux-limited samples was very small the galactic
scale height could not be measured. 

\begin{figure}[t!]
\resizebox{\hsize}{!}{\includegraphics[clip=true]{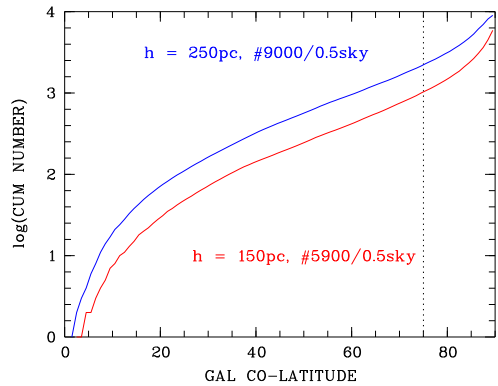}}
\caption{\footnotesize
Cumulative number of non-magnetic cataclysmic variables predicted to be
discovered in the four year eROSITA all-sky survey (eRASS:8). 
}
\label{f:forc}
\end{figure}

At a flux limit of about $10^{-14}$\,\fcgs\ ($0.5-2$\,keV, eRASS:8) one
potentially might discover all CVs brighter than $\log{\rm L_X} = 29, 30, 31, 32$\,
\lxcgs\ within 300, 900, 3000, and 9000\,pc, respectively. The
corresponding volumes are sufficiently large to determine the scale height of
the distribution with high fidelity. In order to estimate the expected number
of CVs in the eRASS, we adopted the parameterized version of the luminosity
function of \cite{pretorius_knigge11}, a truncated power law with index
$-0.8$, and integrated the luminosity function between $10^{28}$\,\lxcgs\ and
$10^{32}$\,\lxcgs.

The results are shown in Fig.~\ref{f:forc} for two adopted values of the
galactic scale height, 150\,pc and 250\,pc, approximating a mildly young
(CVs above the period gap) and a mildly old (CVs below the period gap)
population. The figure shows cumulative numbers as a function of galactic
co-latitude. Hence, one may expect about 6000 and 9000 CVs for the two assumed
values of the scale height in half the sky (a German-Russian agreement gives
exclusive data rights for two hemispheres, respectively, with a dividing line 
along a meridian through the galactic poles and the galactic center). However,
strong absorption in the plane will make CV studies un-feasible at low
latitudes. The expected cumulative numbers at latitude
$|{\rm b^{II}}|> 15$\degr\ are 1000 and 2100, respectively, hence
sufficient to determine the local space density and the galactic scale height
with high confidence, provided the new CVs can be identified as such among the
many other sources.

The number of magnetic CVs, Polars and Intermediate Polars (IPs), that can be
detected in the eROSITA all-sky survey is rather difficult to estimate. Polars
are more abundant in the currently existing samples due to their frequently
present soft X-ray component and the well-matching soft energy response of
ROSAT which led to the discovery of numerous new Polars. 
After ROSAT the detection rate has slowed down,
XMM-Newton made three additions to the class (XGPS-9 just being a candidate) 
\citep{vogel+08, ramsay+09, motch+10}. MCVs on the other hand were discovered
in the last decade serendipitously with RXTE, INTEGRAL and Swift
\citep[e.g.][]{bikmaev+06, butters+07, masetti+06}. Here the IPs are clearly
outstanding due to their comparatively high luminosity at hard X-ray
energies. Their distances are notoriously uncertain. A very simplistic eROSITA IP
forecast can be made by simply scaling the 12 IPs in the XTE all-sky slew
survey (XSS, $3-20$\,keV) with the ratio of the corresponding flux limits 
(the flux in the XSS band and in the hard eROSITA band are the same within a
factor 1.5 for a typical thermal spectrum of $20 - 40$ keV) which reveals
about 15000 IPs. This extrapolation from a more local sample assumes a
constant space density and does not account for the scale height, hence it 
over-predicts the true number considerably. Nevertheless the current 
flux-limited samples of Polars and IPs will grow by orders of magnitude. 

Of particular relevance will be the interplay between eROSITA-discovered CVs
and the distances determined by Gaia which will resolve the distance related
uncertainties in the luminosity functions of non-magnetic and magnetic CVs.


\begin{figure*}[t!]
\begin{center}
\resizebox{0.7\hsize}{!}{\includegraphics[clip=true]{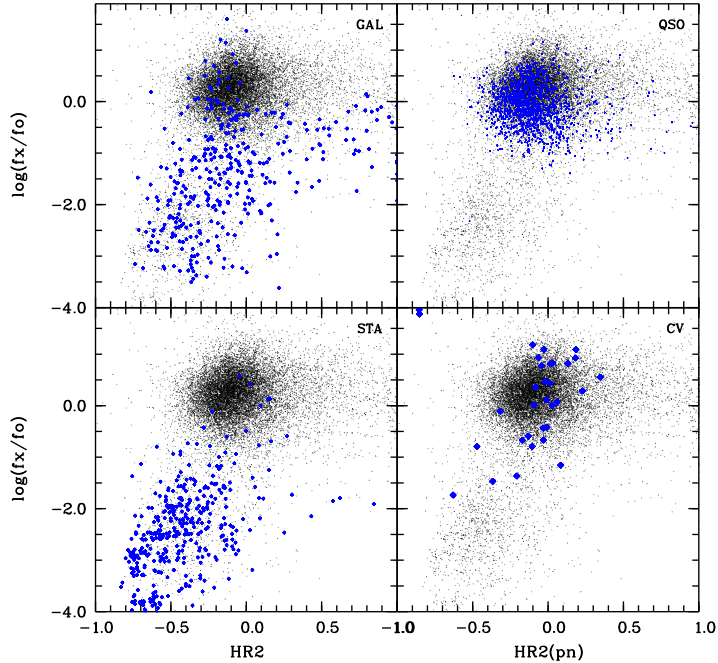}}
\end{center}
\caption{\footnotesize
X-ray to optical flux ratio versus X-ray color of correlated 2XMMi and SDSS
sources. The black background symbols in each panels represent correlated
2XMMi/SDSS  sources \citep{pineauetal11} and blue symbols indicate sources
which were spectroscopically identified in the SDSS.
}
\label{f:lfxo}
\end{figure*}

\section{Source discrimination and optical follow-up}
\begin{figure*}[t!]
\begin{center}
\resizebox{0.7\hsize}{!}{\includegraphics[clip=true]{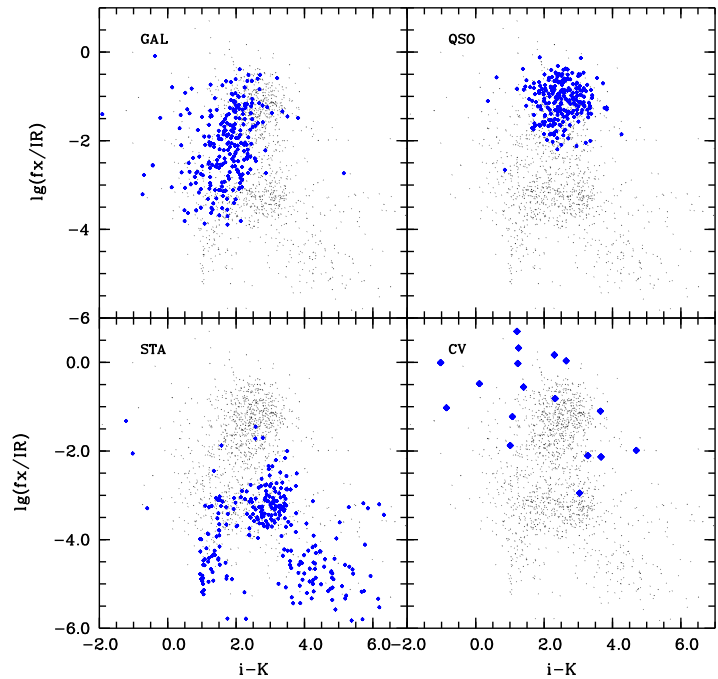}}
\end{center}
\caption{\footnotesize
X-ray to IR flux versus optical to infrared color of abundant X-ray point sources.
}
\label{f:fxir}
\end{figure*}

In Figs.~\ref{f:lfxo} and \ref{f:fxir} we show diagnostic diagrams involving
X-ray, optical and infrared data of point-like X-ray sources detected
serendipitously with XMM-Newton. They may serve as excellent templates and
learning samples for the expected  eROSITA discoveries. 
The X-ray sources were drawn from the 2XMMi-DR3 catalogue
\citep{watson} and were correlated with sources found in SDSS-DR7
\citep{pineauetal11} and with 2MASS. We are adopting the spectral classes
as provided by the SDSS collaboration and do not further separate between different
AGN types and different types of stars. CVs are not classified by the SDSS, 
we are using the correlation with RKCat by \cite{pineauetal11} and correlate
the sample of almost 300 SDSS-CVs compiled by \citet[][and references
  therein]{szkody+11}  as well. The CV sample in Figs.~\ref{f:lfxo} and
\ref{f:fxir} is still rather small and represents a mixture of 
magnetic and non-magnetic objects.  

In Figs.~\ref{f:lfxo} and \ref{f:fxir} we are using the X-ray flux in the
XID-band ($0.5 - 4.5$\,keV). 
The X-ray hardness ratio is defined as HR2 = ($S-$H)/(S+H) where S and H are
the counts in the soft $1.0 - 2.0$\,keV and the hard band $2.0 - 4.5$\,keV,
respectively. The X-ray to optical (infrared) flux ratio was computed as
$\log{f_X/f_{\rm o/IR}} = \log f_X(\mbox{0.5--4.5 keV}) + 0.4 i (K) + 5.37 $. 
The background of objects shown with small black dots is made by objects which
have an associated counterpart in the SDSS photometric catalogue while big
blue symbols indicate a spectroscopic counterpart. 

The QSOs can be rather well located in both diagnostic diagrams. Objects
classified as galaxies will be in most cases low-luminosity AGN with a
considerable contribution from the host galaxy. Stars and Galaxies
have a large overlapping region in Fig.~\ref{f:lfxo}, when infrared data are
included they appear at mutually exclusive regions (Fig.~\ref{f:fxir}). CVs
finally represent the 
least constrained population and have overlaps with QSOs, Galaxies, and Stars
in all parameters. Unless proper motion and parallax data are known for all
involved source classes, significant and clean CV samples cannot be built from
standard data products, i.e.~fluxes from multi-wavelength surveys, that one
finds in current and will find in future archives (e.g.~from Pan-STARRS, DES,
VHS and similar). 

However, plans exist for massive spectroscopic follow-up of eROSITA X-ray
sources which will be purely flux-limited and shall include all types of X-ray
emitters. Projects are underway to build new facilities with high
multiplexity. Perhaps the most ambitious is 4MOST \cite[4-metre Multi-Object
  Spectroscopic Telescope]{dejong+11}, a very large field (goal $>$ 5 square
degrees) multi-object spectrograph with up to 3000 fibers and spectral
resolutions of 5000 and 20000. It is currently in a Conceptual Design phase
and will turn one of the ESO 4-metre telescopes, either VISTA or the NTT, into a
spectroscopy machine that will be able to survey the entire sky accessible
from either La Silla or Paranal. The unique combination of eROSITA detection,
4MOST identification/classification and Gaia distance determination will open
a new 'Golden Age of Cataclysmic Variables'. One will be able to measure 
the contribution of the various CV sub-classes to the Galactic Ridge X-ray
emission and, if orbital periods are available, perform a detailed comparison
with binary population synthesis models \citep{howell+01}.

\begin{acknowledgements}
This project was supported in part by the German DLR under contract 50OR0807.
\end{acknowledgements}

\bibliographystyle{aa}
\bibliography{pal}

\end{document}